\documentclass[procedia]{easychair}

\usepackage{amssymb}
\usepackage{rotating}
\usepackage[section]{placeins}
\usepackage{grffile}

\usepackage{listings}
\usepackage{color}

\usepackage{soul}

\definecolor{dkgreen}{rgb}{0,0.6,0}
\definecolor{gray}{rgb}{0.5,0.5,0.5}
\definecolor{mauve}{rgb}{0.58,0,0.82}

\begin{document}




\title{An automated multiscale ensemble simulation approach for vascular blood flow}
\titlerunning{Multiscale ensemble blood flow}

\author{Mohamed A. Itani\inst{1,2}
\and
Ulf D. Schiller\inst{2}
\and
Sebastian Schmieschek\inst{2}
\and
James Hetherington\inst{2}
\and
Miguel O. Bernabeu\inst{2}
\and
Hoskote Chandrashekar\inst{3}
\and
Fergus Robertson\inst{3}
\and
Peter V. Coveney\inst{2}
\and
Derek Groen\inst{2,4}}

\institute{
Department of Electrical \& Computer Engineering, American University of Beirut, P.O.Box 11-0236, Beirut, Lebanon.
\and 
Centre for Computational Science, University College London, 20 Gordon Street, WC1H 0AJ London, United Kingdom\email{u.schiller@ucl.ac.uk, d.groen@ucl.ac.uk}.
\and
Lysholm Department of Neuroradiology, National Hospital for Neurology and Neurosurgery, University College London, London, United Kingdom.
\and
CoMPLEX, University College London, Physics Building, Gower Street, London, WC1E 6BT, United Kingdom}

\authorrunning{Itani et al.}


\clearpage
\maketitle

\begin{abstract}
Cerebrovascular diseases such as brain aneurysms are a primary cause
of adult disability. The flow dynamics in brain
arteries, both during periods of rest and increased activity, are
known to be a major factor in the risk of aneurysm formation and
rupture. The precise relation is however still an open field of
investigation. We present an automated ensemble simulation method for
modelling cerebrovascular blood flow under a range of flow regimes. By
automatically constructing and performing an ensemble of multiscale
simulations, where we unidirectionally couple a 1D solver with a 3D 
lattice-Boltzmann code, we are able to model the blood flow in a patient artery
over a range of flow regimes. We apply the method to a model of a
middle cerebral artery, and find that this approach helps us to
fine-tune our modelling techniques, and opens up new ways to
investigate cerebrovascular flow properties. 
\end{abstract}




\section{Introduction}

Stroke is a major cause of death and morbidity in the developed world.
Subarachnoid haemorrhage (SAH) is a type of stroke characterised by bleeding
into the fluid around the brain, for example due to the rupture of an 
intracranial aneurysm. An aneurysm is a congenital weakness in a blood
vessel wall which gradually bulges out to form a balloon which can eventually
burst. SAHs represent 5\% of cases of stroke, but is relatively more 
important, as the mortality rate for these events is about 50\%. Overall, approximately 5-10 people per 100,000 are affected
by SAH due to bleeding in the intracranial arterial wall.~\cite{Becske:2010}
The mean age of the victims is 50 years and 10-15\% fail to reach
hospital.  Unruptured aneurysms are much more prevalent, estimated to affect 1-5\% 
of the population of the UK~\cite{nhs-aneurysm}. Indeed, unruptured / asymptomatic 
cerebral aneurysms are a relatively common finding when scanning the brain 
for other reasons~\cite{Becske:2010}. Current methods of determining which aneurysms
have a significant risk of subsequent rupture are based on crude measures such
as aneurysm size and shape, and there is a clear need for a non-invasive tool to
stratify risk more effectively in this large patient group.

Computational fluid dynamics (CFD) techniques may provide means to help quantification of the rupture
risk, if they can incorporate the key conditions affecting brain aneurysms.
Particularly high or low wall shear stress is believed to increase the risk of
aneurysm rupture~\cite{Marques-Sanches:2010}. Researchers increasingly apply
computational fluid dynamics to investigate these
problems~\cite{Grinberg:2011,Formaggia:2003,Fedosov:2014}, and in particular
Shojima et al. concluded that both a very high and a very low  wall shear
stress increases the chance of aneurysm growth and rupture in MCA
aneurysms~\cite{Shojima:2004}. In alignment with these research efforts, we seek 
to establish computational diagnosis and prediction techniques, which may lead to 
major health benefits and reduce the costs of health care in the long term.

An essential driver for these CFD calculations is the flow solver, and
over the last decade several sophisticated and scalable solvers have emerged.
Within this work we rely on HemeLB (described in Section~\ref{Sec:HemeLB}), which
is highly optimized for modelling sparse geometries and has unique optimizations
which allow it to achieve excellent load balance in the presence of complex boundary
and in- and outflow conditions~\cite{Groen:2014}. There are several other
scalable flow solvers that are worth mentioning as well. These include 
the Nektar finite element package~\cite{Nektar,Baek:2009,Grinberg:2012},
the Palabos package~\cite{Palabos,Anzai:2012,Borgdorff:2014}, the Musubi 
environment~\cite{Musubi,Hasert:2013}, MuPhy~\cite{Bernaschi:2009} and WaLBerla~\cite{Gmeiner:2014}.
Although the aforementioned works have provided valuable insight into the
haemodynamic environment of brain aneurysms, little is known about how the
intrinsic variablity of blood flow throughout the day affects aneurysm growth
and rupture. 

The purpose of this paper is to present a tool which automatically creates an
ensemble of multiscale blood flow simulations based on a set of clinically
measurable patient parameters, and runs these simulations using supercomputing
resources.  The tool allows us to automate the study of the blood flow in a
vascular geometry under varying patient-specific conditions. In addition, an
automated data processing component extracts velocity and wall shear stress
(WSS) values, and generates plots and animations which allow us to visualize
these properties in the vascular geometry. This paper builds on previous works
where we simulated flow in arterial networks using a single flow
configuration~\cite{Groen:2013,Bernabeu:2013,Groen:2013-3}.


To showcase our approach, we construct and execute a range of multiscale
simulations of a middle cerebral artery (MCA). We present the results of these
simulations, and compare our approach to related efforts as an initial
validation of our 1D-3D multiscale scheme. This work is
organised as follows. In Section 2, we present the tools we developed to
perform our multiscale ensemble simulations and how we integrate them in an
automated workflow. We describe the setup of our simulation in Section 3, our
results in Section 4 and provide a brief discussion in Section 5.

\section{Automated multiscale ensemble simulations}

Our automated workflow combines three existing components. These
include the HemeLB and pyNS simulation environments, and the FabHemeLB
automation environment.  In this section we describe these three
components, and how they interoperate in our automated multiscale
ensemble simulation (MES) environment.

\subsection{HemeLB}\label{Sec:HemeLB}

HemeLB is a 3 dimensional lattice-Boltzmann simulation environment developed to
simulate fluid flow in complex systems. It is a MPI parallelised C++ code
with world-class scalability for sparse geometries. It can
efficiently model flows in sparse cerebral arteries using up to 32,768
cores~\cite{hemelb,Groen:2013-2} and utilises a weighted domain decomposition
approach to minimize the overhead introduced by compute-intensive boundary and
in-/outflow conditions~\cite{Groen:2014}.  HemeLB allows users to obtain key
flow properties such as velocity, pressure and wall shear at predefined
intervals of time, using a property-extraction framework.

HemeLB has previously been applied to simulate blood flow in healthy brain
vasculature as well as in the presence of brain
aneurysms~\cite{Bernabeu:2013,Mazzeo:2010}. Segmented angiographic data from
patients is read in by the HemeLB Setup Tool, which allows the user to visually
indicate the geometric domain to be simulated. The geometry is then discretized
into a regular unstructured grid, which is used as the simulation domain for
HemeLB. HemeLB supports predefined velocity profiles at the inlets of the
simulation domain, which we generate using pyNS in this work.

\subsection{pyNS: Python Network Solver}

pyNS is a discontinuous Galerkin solver developed in Python, which simulates haemodynamic behaviour in
vascular networks~\cite{pyNS}. pyNS uses aortic blood flow input based
on a set of patient-specific parameters, and combines one-dimensional
wave propagation elements to model arterial vasculature with
zero-dimensional resistance elements to model veins.  The solver
requires two XML files as input data, one with a definition of
the vasculature and one containing the simulation parameters. 
Simulation parameters include mean blood
pressure, cardiac output, blood dynamic viscosity and heart rate.
pyNS has been used in several studies, e.g. to try to 
inform treatment decisions on haemodialysis patients~\cite{Caroli:2013}
and as a large-scale model for distributed multiscale simulations of
cerebral arteries~\cite{Groen:2013}.

\subsection{FabHemeLB}

FabHemeLB is a Python tool which helps automate the 
construction and management of ensemble simulation
workflows. FabHemeLB is an extended version of FabSim~\cite{FabSim}
configured to handle HemeLB operations. Both FabSim and FabHemeLB help 
to automate application deployment, execution and
data analysis on remote resources. FabHemeLB can be
used to compile and build HemeLB on any remote resource, 
to reuse machine-specific configurations, and to organize and curate
simulation data. It can also submit HemeLB jobs to a remote resource
specifying the number of cores and the wall clock time limit for completing
a simulation. The tool is also able to
monitor the queue status on remote resources, fetch results of
completed jobs, and can conveniently combine functionalities into single 
one-line commands. In general, the FabHemeLB commands have the
following structure:
\begin{center}
\begin{verbatim}
fab <target machine> <command>:<parameter>=<value>,...
\end{verbatim}
\end{center}
For example:
\begin{center}
\begin{verbatim}
fab archer ensemble:config=/path/to/config,cores=1536,wall_time=05:00:00
\end{verbatim}
\end{center}

\begin{table}
\caption{List of commands commonly used in FabHemeLB.}
\label{Tab:FabCommands}
\centering
\begin{tabular}{p{2.5cm}|p{10.5cm}}
\hline
command name & brief description\\
\hline
{\tt cold} & Copy HemeLB source to remote resource, compile and build everything.\\
{\tt run\_pyNS} & Execute instances of pyNS to generate a range of flow output files. \\
{\tt generate\_LB} & Convert pyNS output to HemeLB input. \\
{\tt submit\_LB} & Given a set of velocity profiles, submits the corresponding
HemeLB jobs to the remote (supercomputer) resource. \\
{\tt fetch\_results} & Fetch all the simulation results from the
remote resource and save them locally. \\
{\tt analyze} & Performs data-analysis that allows for easy visualization of
the results. \\
{\tt ensemble} & Do all of the above, except {\tt cold}. \\
\end{tabular}
\end{table}

\begin{figure}
\centering
\includegraphics[width=4.0in]{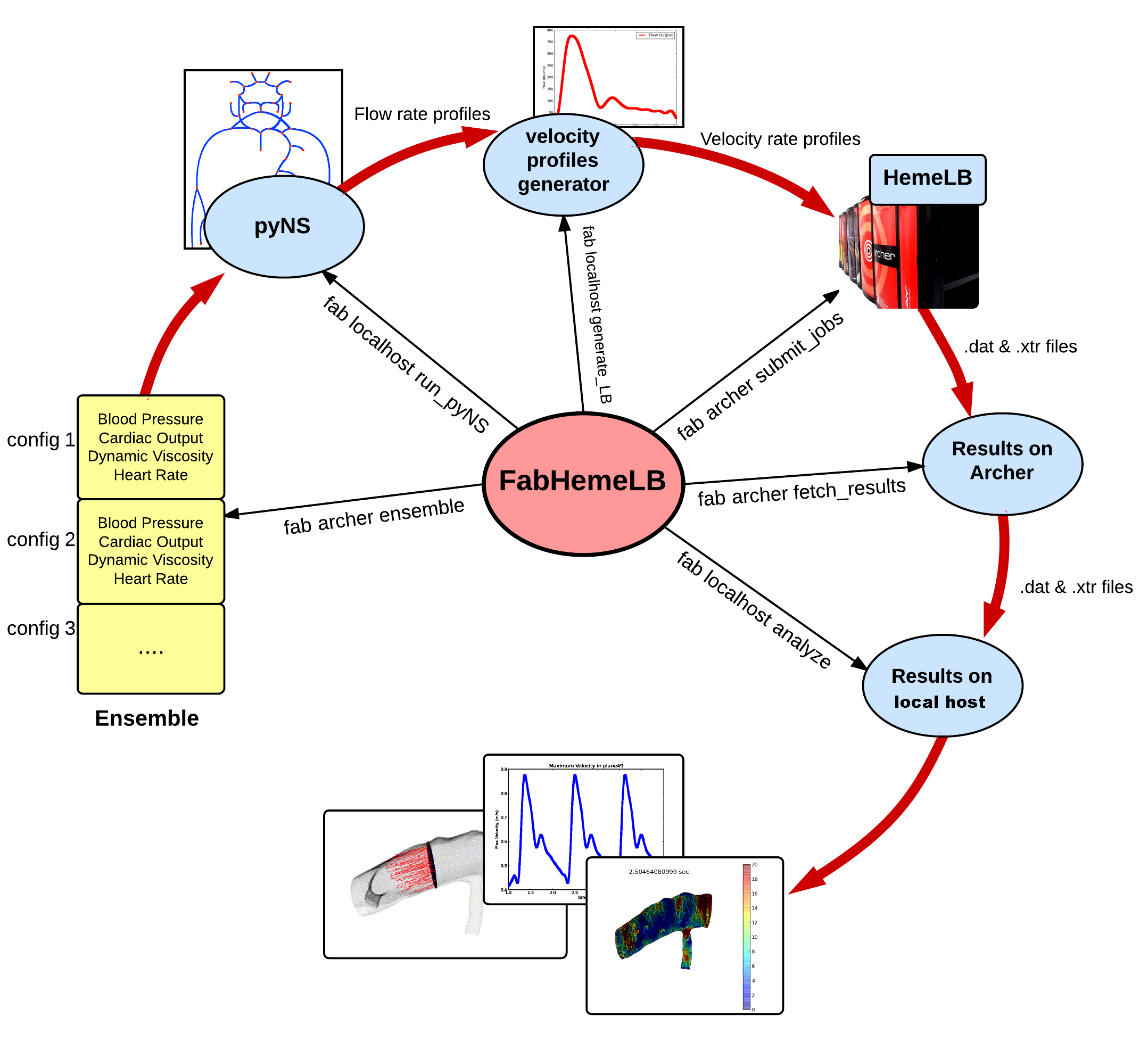}
\caption{Workflow diagram showing the processes involved in the
  ensemble simulation method. The simulations were distributed, with 
PyNS simulations using a local workstation in London, and the HemeLB 
simulations using the ARCHER supercomputer.}
\label{Fig:wrkflw}
\end{figure}

In table~\ref{Tab:FabCommands} we present a number of commands typically executed with FabHemeLB in the scope of this work. 
The commands are customised to run on local machines, continuous integration servers, regional, national or international
supercomputing resources. The workflow is presented in the diagram of Figure~\ref{Fig:wrkflw}. 
Specifically, the {\em analyze} command processes the compressed output files to generate
human readable files and visualisations.  It also generates an image file
showing the whole geometry, wall shear stress within the geometry, and velocity
measurements on pre-selected planes inside the geometry over time as an
animation. 



\section{Setup}

To apply our automated ensemble simulation tool, we employ patient-specific
parameters measured by Sugawara et al. during a study to assess cardiac output
during exercise~\cite{Sugawara:2003}. They measured the blood pressure, cardiac
output and heart rate of 16 young patients (9 male and 7 female) at different
exercise intensities, being at rest or at 70\%, 90\%, 110\% and 130\% of the
ventilatory threshold (VT). The VT is the point during
exercise training at which pulmonary ventilation becomes disproportionately
high with respect to oxygen consumption. This is believed to reflect onset of
anaerobiosis and lactate accumulation.  We add two more sets of values at
30\% and 50\% VT, linearly extrapolating the other parameters. The resulting
seven sets of parameters are used in our automated workflow, resulting in
seven HemeLB simulations being run. We present the seven sets of parameters 
in Table~\ref{Tab:configs}. 


\begin{table}[ht!] 
\caption{Configuration data used as input for pyNS to run the ensemble
  of simulations. The values are based on the average of the measurements 
  of 16 people at different exercise
  intensities measured by the percentage of the ventilatory
  threshold(VT)~\cite{Sugawara:2003}. In the last column, we provide the mean 
  flow velocity in the right MCA, as calculated using PyNS, for each configuration.}
\label{Tab:configs}
\centering 
\begin{tabular}{llllll} 
\hline 
Configuration & Exercise  & Blood Pressure & Cardiac & Heart & mean flow\\
              & intensity & Mean           & output & rate & velocity\\
&  & [mmHg] & [L/min] & [bpm] & [$ms^{-1}$]\\
\hline
1  & Rest  & 80 & 4.8 & 68 & 0.460\\
\hline
2 & 30\% VT  & 87 & 6.2 & 79 & 0.451\\
\hline
3 & 50\% VT  & 94 & 7.6 & 90 & 0.428\\
\hline
4 & 70\% VT  & 100 & 9 & 101 & 0.393\\
\hline
5 & 90\% VT  & 112 & 10.7 & 113 & 0.371\\
\hline
6 & 110\% VT  & 116 & 11.9 & 120 & 0.351\\
\hline
7 & 130\% VT  & 122 & 13.2 & 134 & 0.339\\
\end{tabular}
\end{table}

\begin{figure}
\centering
\includegraphics[width=3.0in]{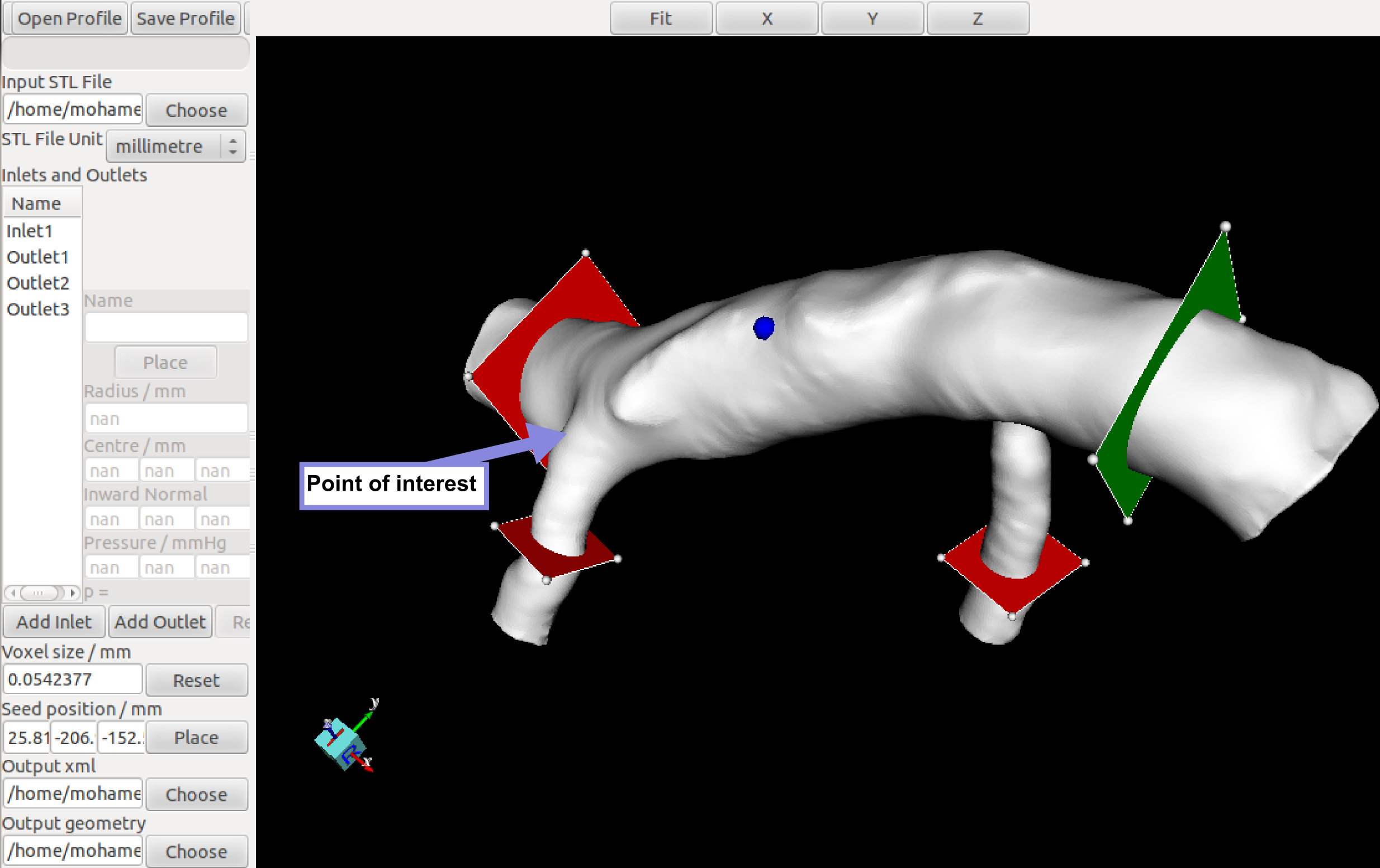}
\caption{HemeLB Setup Tool loaded with a middle cerebral artery
  geometry. The inlet is indicated by a green surface, and the three
outlets with a red surface. The lattice site which we used for in-depth
wall-shear stress analysis is highlighted with a light blue arrow.}
\label{Fig:setup-tool}
\end{figure}

For our pyNS simulations, which we ran for 10 cardiac cycles, we set the blood
density to 1050 $Kg/m^{3}$, Poisson's ratio of transverse to axial strain to
0.5 and the time step to 5 ms. For our HemeLB runs we use a model derived from
a patient-specific angiographic 3D geometry of a middle cerebral artery,
supplied by the Lysholm Department of Neuroradiology, University College
Hospital, London and segmented using the GIMIAS tool~\cite{Gimias}.  We use a
voxel size of $18.9\mu m$, which results in a geometry containing 13,179,961
lattice sites. In Figure~\ref{Fig:setup-tool} we show the setup tool interface
with the MCA geometry. Here the inlet is given by the green plane and the
outlets by the red planes. The location of interest for our WSS analysis is highlighted.
For simulations of this particular voxelized simulation domain,
we specify a time-step of $0.5014 \mu s$ and run each simulation for 7.9
million steps, or 4 seconds of simulated time.

During our HemeLB runs, we store the WSS throughout the geometry for every
50,000 time steps. In addition, we define an output plane close to the outlet,
at 49mm from the ear, 2mm away from the outflow boundary (or {\em outlet}), to
record velocity and pressure data every 50,000 steps.  In all our runs we use
interpolated Bouzidi wall boundary conditions~\cite{Bouzidi:2001} and
zero-pressure outlet conditions, (for details see~\cite{hemelb}). Using the
{\it ensemble} command, we run an ensemble of 7 simulations on the ARCHER
supercomputer. In total, we used 10,752 (7 $\times$ 1536) cores for a duration
of approximately 3.5 hours. 

\section{Results}

\begin{figure}
\centering
\includegraphics[width=2.3in]{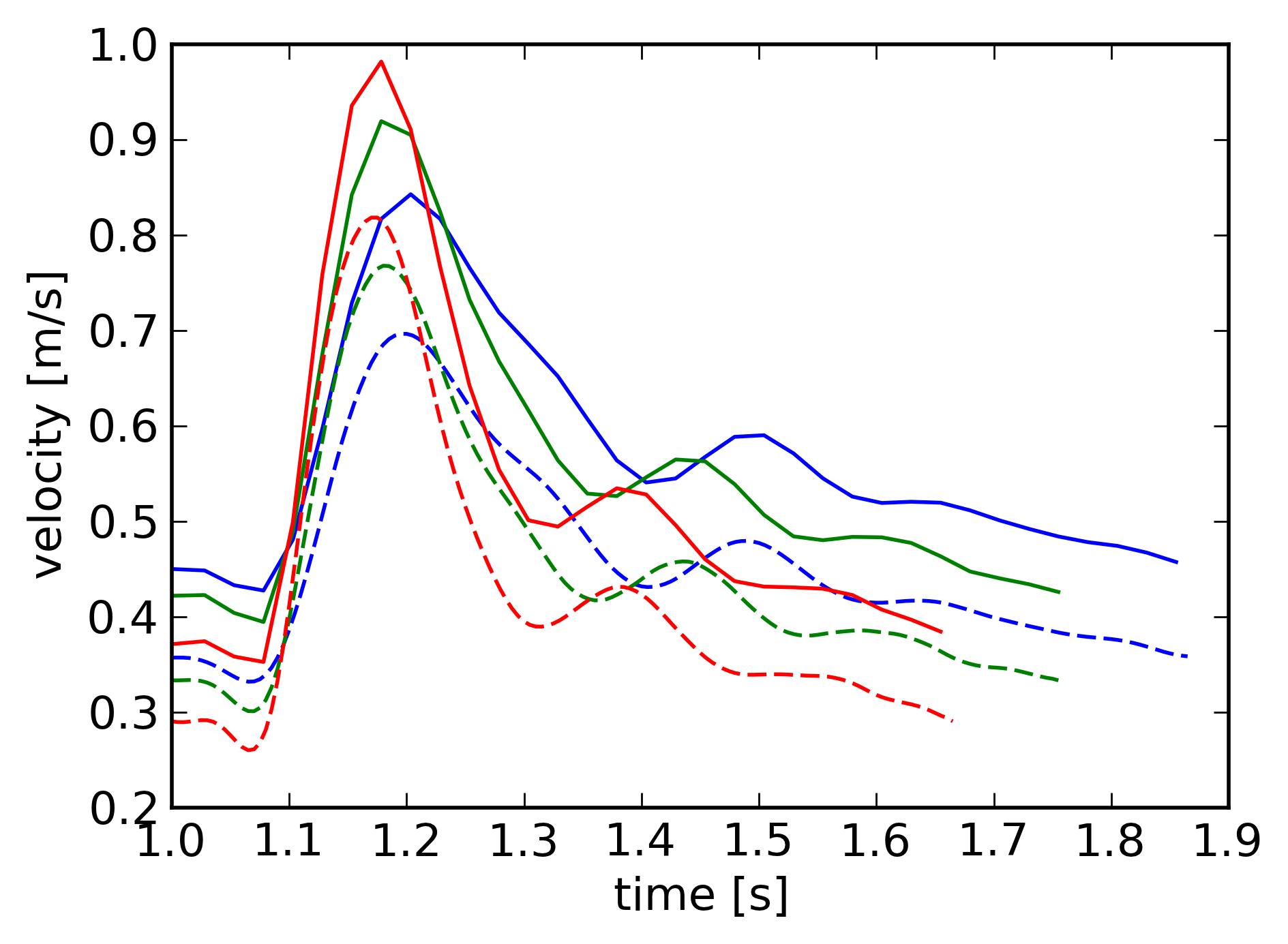}

\caption{Velocity time-series for the configurations 1 (bottom lines)
  to 3 (top lines). Dashed lines indicate the velocity time-series
  generated for the inflow boundary by pyNS, and solid lines correspond to the maximum
  velocity measured at the 49mm plane in HemeLB.}

\label{Fig:1-3}
\end{figure}

\begin{figure}
\centering
\includegraphics[width=2.3in]{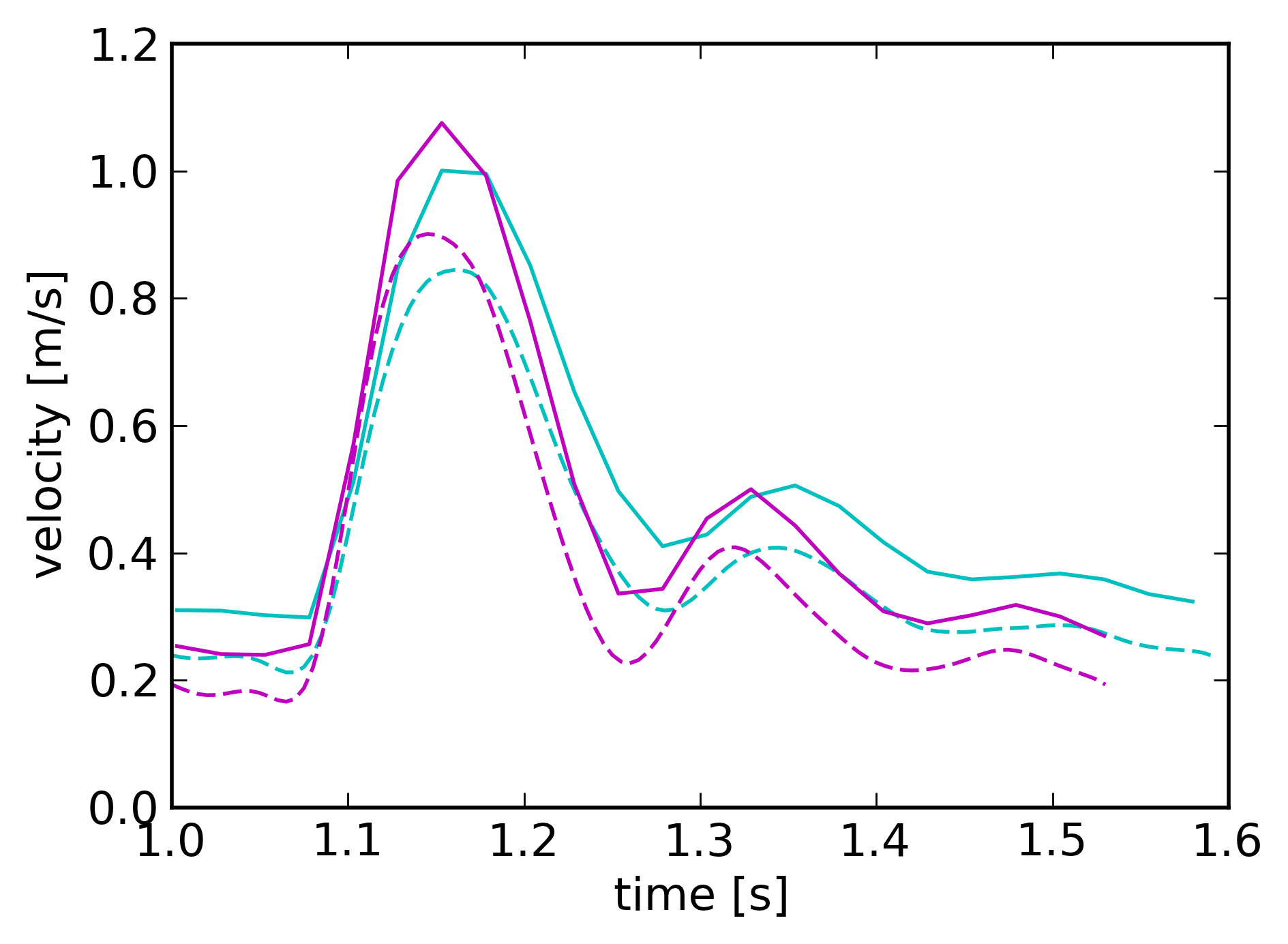}
\includegraphics[width=2.3in]{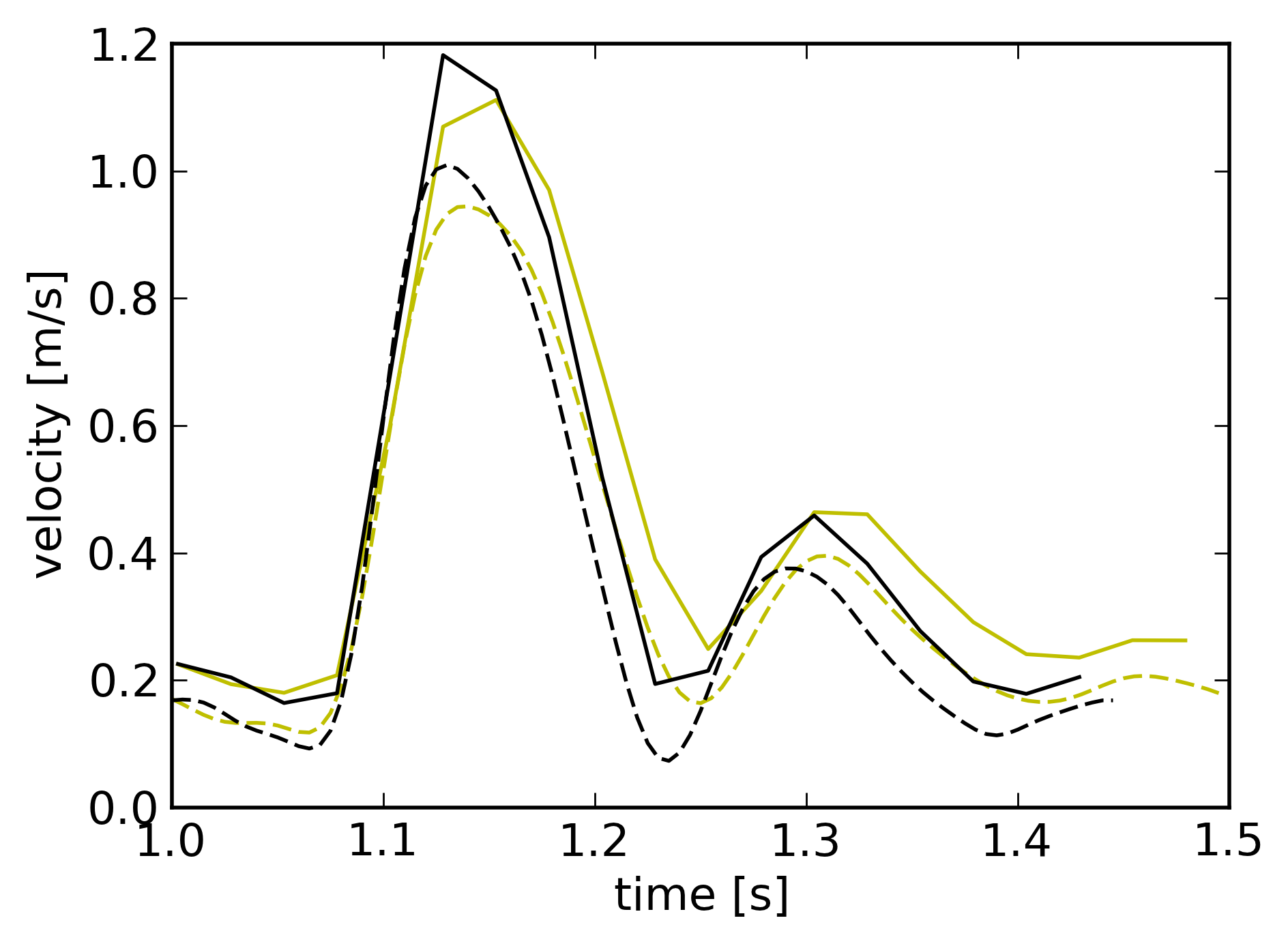}

\caption{Left: Velocity time-series as in Fig.~\ref{Fig:1-3} but for
  configurations 4 (bottom line) and 5 (top line). Right: Velocity time-series 
as in Fig.~\ref{Fig:1-3} but for configurations 6 (bottom line) and 7 (top line).}

\label{Fig:4-7}
\end{figure}

We present the time series of the maximum velocity in the measurement plane in Figures \ref{Fig:1-3} and \ref{Fig:4-7}
for the parameters listed in Table \ref{Tab:configs}. The curves demonstrate
that the frequency of the 1D cardiac cycles generated by pyNS are accurately
reproduced in the 3D high-resolution HemeLB simulations. The peak velocity in
the measurement plane is higher than the input velocity because the measurement
plane has a lower area but the total flux remains constant throughout the MCA
due to the incompressible flow.

Within the ensemble, the frequency of the cardiac cycles increases
with exercise intensity as the heart rate increases. While this seems
trivial, it indicates that the time-resolution chosen to couple pyNS
and HemeLB is sufficient to reproduce this effect. The good match
between the input and the measurement gives confidence that the LB
simulations have converged and the cardiac cycles are stable.

\subsection{Wall shear stress}

\begin{figure}
\centering
\includegraphics[width=4.0in]{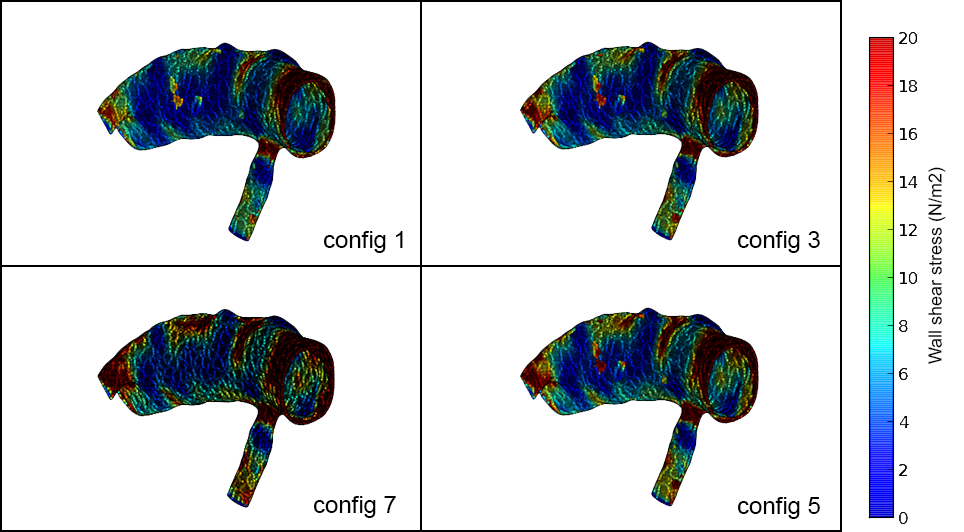}
\caption{A snapshot of the wall shear stress for the configurations 1,
  3, 5 and 7 at the peak systole. Red regions indicate high wall shear
  stress while blue indicates low wall shear stress. Exercise
  intensity is increasing clockwise for the configurations
  shown.}
\label{Fig:stress}
\end{figure}

Figure~\ref{Fig:stress} shows snapshots of the WSS for four configurations of
the ensemble at the peak systole, when the flow velocity is highest. The colour
scale allows us to identify regions of high WSS which are located
predominantly at constrictions of the MCA and around the inlet in an
assymetric pattern. Based on the WSS values near the inlet, we conclude that
the standard circular-shaped Poisseuille velocity profile used in HemeLB is ill
suited for patient-specific geometries, as they usually feature non-circular
inlets. As a result, we are now developing a method for a modified velocity
profile which takes non-circular inlet shapes into account.

\begin{figure}
\centering
\includegraphics[width=2in]{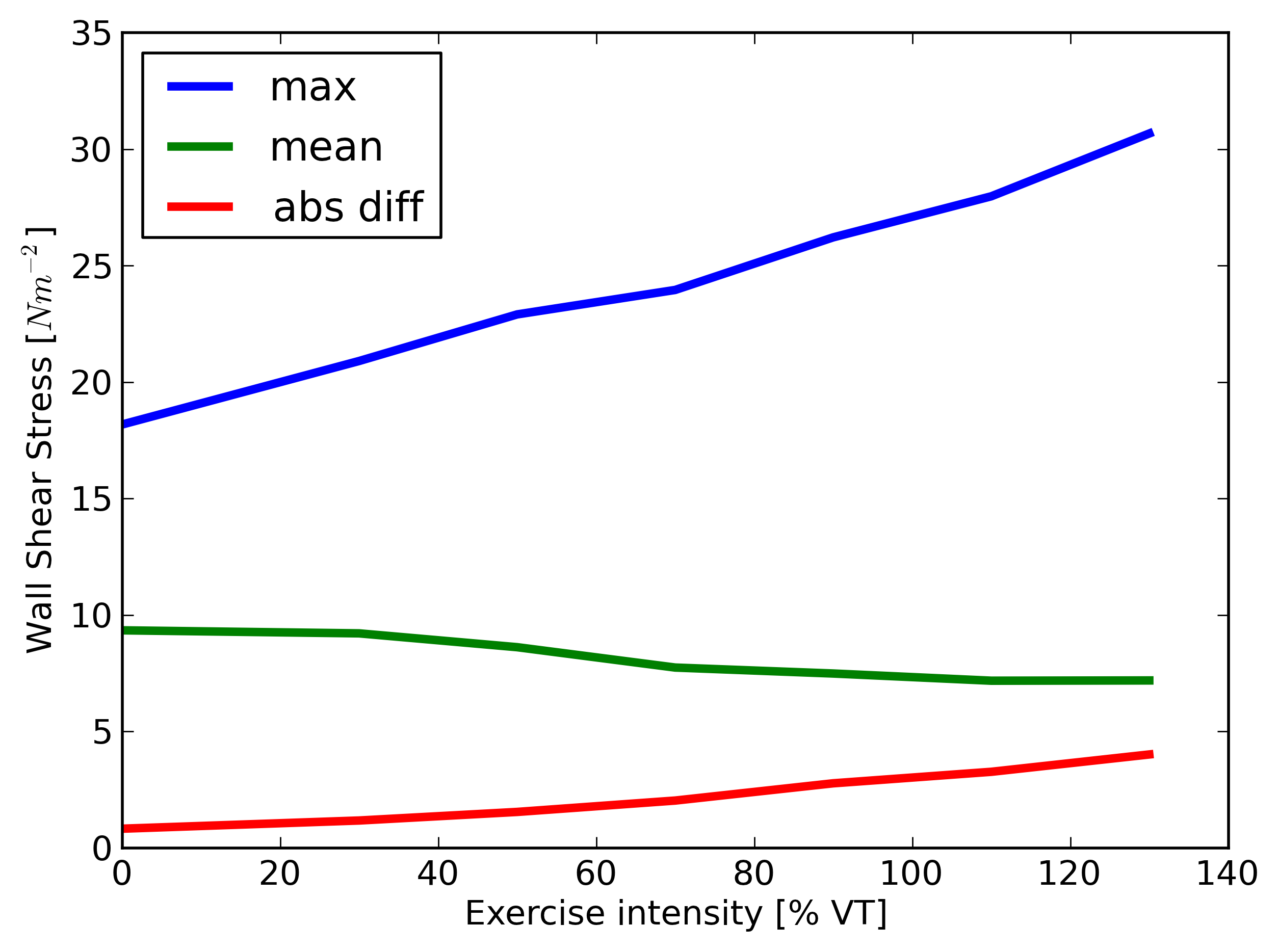}
\caption{Cross-instance analysis of the WSS for the location of interest indicated in Fig.~\ref{Fig:setup-tool}. We 
extracted WSS values in the simulations for a range of 3 cardiac cycles 
(when at rest) up to 6 cardiac cycles (at 130\% VT). We present the maximum measured WSS (blue line), the average WSS (green), and
the time average of the WSS slope, calculated over intervals of 0.025 $s$ (red). The
WSS values at rest can be found on the left side (0\% VT).}
\label{Fig:stress-local}
\end{figure}

At the point of interest indicated in Fig.~\ref{Fig:setup-tool}, we observe
much higher wall shear stress at higher exercise intensity. We present a
cross-instance analysis of the WSS at this point in
Fig.~\ref{Fig:stress-local}. Here we provide for instance the mean WSS, which
decreases linearly with the mean velocity at the inlet (see
Table~\ref{Tab:configs}, both values are reduced to $\sim$0.75 times the
magnitude at rest, when measured at 130\% VT exercise intensity). This matches the
theoretically expected linear scaling of the WSS with the velocity parallel to the wall.
This parallel velocity is expected to be proportional to the velocity at the
inlet for simple geometries and flow regimes. Our WSS results are in
line with related literature, which report maximum values in MCAs in the range
of 14 to 40 $Nm^{-2}$ for unruptured aneurysm
geometries~\cite{Shojima:2004,Cebral:2011}.

We also present the maximum WSS in time, which is $\sim 18 Nm^{-2}$ for the run
at rest, with a heart rate of 68 bpm and a maximum velocity of $0.84 ms^{-1}$.
At full exercise intensity, the maximum WSS is much higher, at $\sim 31
Nm^{-2}$ for a heart rate of 134 bpm and a maximum velocity of $1.19 ms^{-1}$.
Here, while the mean WSS and velocity decrease with exercise intensity, the
maximum WSS and velocity increase. Between these two cases, the difference in
maximum WSS (a factor of 1.77) cannot be justified solely by the difference in
maximum velocity (a factor of 1.45). Indeed, the (much larger) difference in
heart rate (a factor of 1.97) may be an important contributor to the magnitude
of the maximum WSS. Further investigations are required to explore the exact
nature of these relations. The variability of the WSS, here measured as the
averaged absolute difference between consecutive extractions at a 0.025 s
interval, increases as expected with higher exercise intensity.  


\section{Discussion and conclusions}

We present an automated ensemble simulation framework and its
application to model blood flow in the middle cerebral artery under a range
of patient-specific cardiac parameters, using a multiscale ensemble approach. 
We show good agreement of velocity profiles at the inlet with those close
to the outlet, and that our non-lattice aligned inflow conditions require 
further enhancement.
FabHemeLB allows us to run the whole workflow for the
relatively complicated setup in one tool, including the execution and analysis
of the ensemble simulations. It reduces the human effort required for doing
these tasks, and by automatically scheduling the ensemble
instances in parallel it also allows for efficient use of large core counts and
a reduced time to solution. The systematic execution and analysis patterns 
offered by FabHemeLB allow us to easily identify shortcomings in our existing 
approach. Not only does this feature in FabHemeLB boost our ongoing research, 
it also provides the level of data curation required to do future, more extensive, 
validation studies.

In our case study, we investigate the wall shear stress (WSS) properties
in a middle cerebral artery at a location of interest close to the outlet. We
find that the mean WSS correlates as expected linearly with the average
flow velocity at the inlet. However, in addition we find evidence that the maximum WSS is
dependent on the heart rate as well as the average flow velocity. This implies
that these relations are non-trivial, and that a comprehensive analysis of flow
dynamics in cerebral arteries should not only include the presence of pulsatile
flow, but also the presence of these flows over a range of heart rates.

\section{Acknowledgements}

We are grateful to Rupert Nash for his efforts on enabling property extraction
for HemeLB, and to Aditya Jitta for performing the segmentation. This work has
received funding from the CRESTA project within the EC-FP7 (ICT-2011.9.13)
under Grant Agreements no. 287703, and from EPSRC Grants EP/I017909/1
(www.2020science.net) and EP/I034602/1. This work made use of the ARCHER
supercomputer at EPCC in Edinburgh, via EPSRC and the UK Consortium on
Mesoscopic Engineering Sciences (EP/L00030X/1).  We have also used the
Oppenheimer cluster, administered at the Chemistry Department at University
College London.


\bibliographystyle{unsrt}
\bibliography{ensemble}







\end{document}